\documentclass{aa}
\usepackage{graphicx}
\usepackage{txfonts}
\usepackage{natbib}
\begin{document}
\title{Technetium and the third dredge up in AGB stars}
\subtitle{II. Bulge stars\thanks{Based on observations at the
Very Large Telescope of the European Southern Observatory,
Cerro Paranal/Chile under Program 65.L-0317(A, B).}}

\author{S. Uttenthaler\inst{1}\fnmsep\inst{2}
\and
J. Hron\inst{2}
\and
T. Lebzelter\inst{2}
\and
M. Busso\inst{3}
\and
M. Schultheis\inst{4}
\and
H. U. K\"aufl\inst{1}
}

\offprints{S. Uttenthaler, suttenth@eso.org}

\institute{European Southern Observatory, Karl Schwarzschild Stra\ss e 2,
85748 Garching near Munich,
Germany\\
\email{(suttenth;hukaufl)@eso.org}
\and
Department of Astronomy, University of Vienna, T\"ur\-ken\-schanz\-stra\ss e
17, 1180 Vienna, Austria\\
\email{(hron;lebzelter)@astro.univie.ac.at}
\and
Department of Physics, University of Perugia, Via A. Pascoli 1,
06123 Perugia, Italy\\
\email{busso@fisica.unipg.it}
\and
Observatoire de Besan\c{c}on, 41bis, avenue de l'Observatoire, BP 1615,
25010 Besan\c{c}on Cedex, France\\
\email{mathias@obs-besancon.fr}
}

\date{Received April 20, 2006; accepted}


\abstract
{We searched for Technetium (Tc) in a sample of bright oxygen-rich
asymptotic giant branch (AGB) stars located in the outer galactic
bulge. Tc is an unstable element synthesised via the s-process in
deep layers of AGB stars, thus it is a reliable indicator of both
recent s-process activity and third dredge-up.}
{
We aim to test theoretical predictions on the luminosity limit for
the onset of third dredge-up.}
{Using high resolution optical spectra obtained with the UVES
spectrograph at ESO's VLT we search for resonance lines of neutral
Tc in the blue spectral region of our sample stars. These measurements
allow us to improve the procedure of classification of stars with
respect to their Tc content by using flux ratios. Synthetic spectra
based on MARCS atmospheric models are presented and compared to the
observed spectra around three lines of Tc. Bolometric magnitudes are
calculated based on near infrared photometry of the objects.}
{Among the sample of 27 long period bulge variables four were found to
definitely contain Tc in their atmospheres.}
{The luminosity of the Tc rich stars is in agreement with predictions
from AGB evolutionary models on the minimum luminosity at the time when
third dredge-up sets in. However, AGB evolutionary models and a
bulge consisting of a single old population cannot be brought into
agreement. This probably means that a younger population is present in
the bulge, as suggested by various authors, which contains the Tc-rich
stars here identified.
}


\maketitle
%

\section{Introduction}

The Asymptotic Giant Branch (AGB) phase is an important step in the
late evolution of low and intermediate mass (1 - 8~M$_{\sun}$)
stars. In the most luminous part of the AGB the behaviour of a star is
characterised by the so called Thermal Pulses (TP), thermal
instabilities of the He shell accompanied by changes in luminosity,
temperature, period and internal structure \citep{Busso99,Herwig}.
Between the repeated events of explosive He-burning,
heavy elements can be produced via the ``slow neutron capture
process''  \citep[s-process, see e.g.][]{Wallerstein97} in the region
between the hydrogen and the helium burning shells. The processed
material is then brought to the stellar surface by the convective
envelope that temporarily extends to these very deep layers. This
mixing event is called the third dredge-up (3DUP) and is the cause
of the eventual metamorphosis of an oxygen-rich M-star to a
carbon-rich C-star.

In recent years, considerable progress has been made with regard
to models for the 3DUP and nucleosynthesis on the thermally pulsing
AGB \citep[TP-AGB;][and references therein]{Busso99,Lugaro03}. The
different evolution models agree qualitatively in the sense that the
3DUP is more efficient for more massive convective envelopes
\citep[e.g.~][]{Straniero97} and for lower metallicities
\citep[e.g.~][]{Straniero03}. However, the quantitative results are
still model dependent \citep{Lattanzio02,Lugaro03} and grids of new
models covering a wider range of stellar parameters are scarce.
In general, observed abundances of s-process elements agree
with the model predictions at least qualitatively. For instance,
the metallicity dependence of the 3DUP is supported by observations
\citep{Busso01,Abia02}, although observations not compatible with
the models have to be mentioned here \citep{Deroo,Masseron}. A
first attempt to directly check the conditions for the onset of 3DUP
observationally has been made by \citet{LebzelterHron}. Important
constraints on the minimum (core) mass (and hence luminosity) for
3DUP and its efficiency also come from the observed luminosity
function of carbon stars in the LMC \citep{CostaFrogel} and in the
Galaxy \citep{Guandalini2006}, together with synthetic stellar
evolution calculations \citep{GroenewegendeJong,Marigo96}.

Tech\-net\-ium (Tc) is among the elements produced by the s-process and
has no stable isotopes. The isotope of Tc with the longest half-life time
produced via the s-process is \element[][99]{Tc} with
$\tau_{1/2} = 2.1 \times 10^{5}$ years. In the following discussion we
are always referring to the isotope \element[][99]{Tc} when we discuss
Tc. The short half-life time makes Tc a reliable indicator of the 3DUP,
because any Tc we see in a star must have been produced during its
previous evolution on the TP-AGB. Tc should be detectable at the surface
after only a few thermal pulses \citep{Goriely}. It should be noted at
this point that the absence of Tc does not necessarily mean the absence
of TPs but rather the absence of 3DUP for several TPs. This could be
caused by an initial mass on the TP-AGB which is too low or by a mass
loss rate at the end of the AGB-evolution which is too high
\citep[e.g., ][]{Busso92}.

A number of studies have been published on observations of Tc in
spectra of late type stars: Starting from the first observation by
\citet{Merrill}, via investigations by \citet{DominyWallerstein86},
\citet{WallersteinDominy88}, \citet{SmithLambert88,SmithLambert90},
\citet{Vanture91}, to studies on S-stars \citep{VanEckJorissen99}
and C-stars \citep{BarnbaumMorris93}. Important studies on Tc in
large samples of long period variables have been conducted by
\citet{Little79}, \citet{Little87} and \citet{LebzelterHron,paperI}.

%
%

In the first paper of this series \citep[][ hereafter called paper~I]{paperI},
the Tc content of a sample of luminosity selected galactic field AGB stars
was studied. A significant number of stars above the luminosity limit for
3DUP, indicated by mixing models, were found to not show Tc in their spectra.
This can be explained by the fact that the second important parameter for
3DUP is the mass of the envelope. It is suspected that the absence of Tc in
a significant fraction of long period Miras is due to a reduction of the
envelope mass below the critical limit by mass loss and/or due to a low
initial mass of the star. Due to the uncertainties in distance (based on
Hipparcos parallaxes) of field stars, no definite conclusions could be drawn.

To improve the situation, a sample of targets with more accurate distances
is required. Given the current accuracy of distance measurements of AGB
stars and the low flux in the blue spectral region of these stars, the
only available targets for such studies can be found in the galactic bulge.
The distance to the bulge (8~kpc) is known rather accurately and the depth
of the bulge is low enough at least in the outer parts to have a fairly
low depth induced scatter in brightness
\citep[$+0\fm5 / -0\fm6$, ][]{Schulthe98}. Using ESO's VLT, exposure times
are short enough to execute observations of a statistically relevant sample
in a reasonable time. Additionally, the bulge population is expected to be
more homogeneous than the disk population: The fact of widely absent C-stars
indicates that high mass stars are no longer present in the bulge. We
therefore chose to observe bright AGB stars in the Palomar Groningen field
no.~3.

It should be noted, however, that AGB variables in the bulge do not have the
same average properties as their disk counterparts, especially the Semiregular
Variables (SRVs). These differences might be explained by a different
age-metallicity relation and a different pulsation mode for the bulge SRVs
compared to the field SRVs.
For details we refer to \citet{Schulthe98}, while the present paper focuses on
dredge-up indicators rather than pulsation properties.

The paper is structured in the following way: The sample selection is
presented in section~2, possible foreground and background contamination is
discussed in section~3. In section~4 the UVES observations are
presented, while sections~5, 6 and 7 deal with the basic characteristics of
the sample stars summarised in Table~\ref{charact}, the detection of Tc and
the discussion of the data, respectively. Conclusions can be found in
section~8.


\section{Sample selection}

The selection of the sample was limited to oxygen-rich long period variables in
the Palomar Groningen field no.~3 (PG3) which is centred 10\degr\
south of the galactic centre and covers an area of 6.\degr5 $\times$
6.\degr5 on the sky. It is located in the outer bulge where
interstellar extinction is rather low and the depth of the bulge is
small. The PG3 field has been studied extensively in the past
\citep{Wess87,Blommaert,Ng,SchultheDiss,Schulthe98}
and periods for a considerable number of variable stars are known.

In order to avoid foreground or background objects, we constructed a period
K-magnitude diagram based on near-infrared photometry acquired at the ESO 1-m
telescope at La Silla \citep[see ][ and references therein]{Schulthe98} and
additional data from DENIS \citep{Epchtein97} for some stars. The periods
were taken from \citet{Wess87}. A range of $\pm1\fm0$~mag around the period
K-magnitude relation for the SgrI field from \citet{Glass95} 
was allowed for the potential targets to account for the depth of
the bulge and the intrinsic scatter in brightness (partly only single epoch
measurements available). Stars outside this range were considered to be in
the foreground or background. The targets were chosen
to be brighter than the RGB-tip (8.2 mag in K$_0$ at the bulge distance, see
Tiede, Frogel \& Terndrup, 1995; Omont et al., 1999) to sample the AGB
up-wards. The distribution was chosen to be about equal between Semiregular
and Mira variables. This procedure resulted in a preliminary target list for
the WFI observations.

\begin{table*}
\caption{Basic characteristics of the targets in our sample. Column~1 lists the
stellar identifier, adopted from \citet{Wess87} which codes the variability
type: M stands for Mira variable, S for Semiregular variable. Columns~2 and 3
give the J2000 coordinates as found in the 2MASS catalogue. The adopted value
of the period is given in column~4. Columns~5 and 6 give the mean J and K
magnitudes derived from DENIS, 2MASS, and own measurements at the ESO
La Silla 1-m telescope. They are corrected for interstellar extinction using
the laws of \citet{Schulthe98} and \citet{GS2003}. Column~7 is the radial
velocity as determined from the UVES spectra. Finally, column~8 lists whether
Tc could be detected in the spectrum of the respective object. For details
see section 5.}\label{charact}
\begin{tabular}{lccrrrrc}
\hline\hline
Name           & RA          & Dec        &  Period [d] & $J_0$ [mag] & $K_0$ [mag] & RV [kms$^{-1}$] & Tc? \\
\hline			   			     	              		      		        
\object{M45}   & 18 12 48.5  & -33 19 27  & 271.02      &  8.19       &  6.69       & +19.8           & no  \\
\object{M100}  & 18 13 38.5  & -36 40 03  & 298.7       &  7.65       &  6.37       & -37.3           & no  \\
\object{M143}  & 18 14 13.6  & -32 36 58  & 204.19      &  9.32       &  8.00       & +17.6           & no  \\
\object{M195}  & 18 15 07.7  & -33 09 22  & 216.59      &  8.88       &  7.61       & -142.8          & no  \\
\object{M277}  & 18 16 15.4  & -31 42 49  & 263.23      &  8.55       &  7.16       & -34.5           & no  \\
\object{M315}  & 18 16 45.0  & -32 40 01  & 326.8       &  8.02       &  6.58       & -61.9           & no  \\
\object{M331}  & 18 16 57.6  & -32 06 29  & 311.07      &  8.17       &  6.64       & -31.8           & no  \\
\object{M626}  & 18 21 32.5  & -36 07 35  & 298.48      &  8.44       &  7.19       & -67.7           & yes \\
\object{M794}  & 18 24 28.0  & -32 30 51  & 303.54      &  7.48       &  6.07       & -51.0           & no  \\
\object{M1147} & 18 33 06.2  & -36 22 27  & 395.63      &  7.23       &  5.76       & +1.0            & yes \\
\object{M1179} & 18 33 54.7  & -35 01 19  & 274.51      &  8.56       &  7.20       & +61.8           & no  \\
\object{M1287} & 18 36 44.5  & -32 25 43  & 312.5       &  8.03       &  6.67       & +229.4          & no  \\
\object{M1313} & 18 37 35.5  & -34 12 27  & 378.7       &  8.42       &  6.79       & +14.1           & no  \\
\object{M1347} & 18 38 45.7  & -34 33 28  & 426.6       &  7.47       &  5.99       & +60.5           & yes \\
\object{S70}   & 18 13 05.6  & -32 23 41  & 166.52      &  8.56       &  7.19       & -77.2           & no  \\
\object{S328}  & 18 16 56.2  & -33 53 24  & 161.3       &  9.01       &  7.78       & -117.9          & no  \\
\object{S639}  & 18 21 42.3  & -35 03 08  & 167.3       &  8.74       &  7.46       & -88.7           & no  \\
\object{S719}  & 18 23 18.4  & -36 05 05  & 279.77      &  7.90       &  6.61       & +69.4           & no  \\
\object{S942}  & 18 28 09.0  & -36 31 26  & 338.0       &  7.86       &  6.54       & -77.8           & yes \\
\object{S1002} & 18 29 22.8  & -31 44 16  & 194.23      &  8.41       &  7.11       & +33.0           & no  \\
\object{S1008} & 18 29 34.1  & -34 08 27  & 232.14      &  8.00       &  6.73       & -39.0           & no  \\
\object{S1059} & 18 30 42.8  & -36 00 36  & 144.1       &  8.91       &  7.68       & +18.0           & no  \\
\object{S1176} & 18 33 43.8  & -31 27 35  & 184.1       &  8.38       &  6.99       & +56.1           & no  \\
\object{S1204} & 18 34 38.8  & -34 00 08  & 197.0       &  8.67       &  7.32       & +236.5          & no  \\
\object{S1470} & 18 42 31.7  & -35 59 28  & 184.08      &  8.51       &  7.33       & -37.7           & no  \\
\object{S1517} & 18 27 19.1  & -32 06 33  & 188.8       &  9.03       &  7.71       & +80.2           & no  \\
\object{S1991} & 18 15 31.4  & -31 58 40  & 124.7       &  9.28       &  8.06       & -102.1          & no  \\
\hline
\end{tabular}
\end{table*}

\section{Fore- and background contamination}

Foreground confusion is a serious problem when observing in the
direction of the bulge. The density of stars drops rather rapidly
behind the bulge, and a star would have to be extremely bright if
it was to fall on the period luminosity relation. Thus background
contamination can be considered as low and we can restrict the
discussion to foreground contamination. 

Since long period variables obey period luminosity relation(s),
their distance can be inferred from their position in such a
diagram with a certain accuracy. Though, for single stars
membership in the disk can not be excluded if no kinematic
information (proper motion) is available. Near-by M dwarfs in
the galactic disk and stars ascending the RGB can be excluded
from our sample since neither fulfils the variability
criteria of long period variables.

To get an estimate of the possible foreground contamination we used
the Besan\c{c}on model of population synthesis \citep{Robin03}. As a
representative field, we calculated the population of the central
square degree of the PG3 field located at $b=-10\degr$ and $l=0\degr$
using the observed photometric ranges (apparent K magnitude, $(J - K)_{0}$
colour) as criteria (the extinction towards the PG3 field is rather
low, see Sect.~\ref{SectResults}). No criteria for the pulsation could be
included, but stars with too low bolometric magnitude were treated as
non-variable. The result indicates foreground AGB contamination at
a level of 2.4\%, which gives on average less then 1 star in a sample of
27 (the final number of objects).
Thus, we are confident we have only bulge stars in our sample.

\section{Observations}
\label{SectObs}

%


From the preliminary target list, 27 objects were selected for the UVES
observations. Details on this high resolution optical spectrograph can be
found in \citet{Dekker}. Pulsation phases of the targets were roughly
estimated from preceding B-band Wide Field Imager (WFI) observations
obtained at the ESO/MPG 2.2m telescope during two runs in April and
May 2000. These measurements were used for the final sample selection in
an attempt to observe only targets with their maximum brightness close to
the time of the UVES observations. The final object selection was also
strongly driven by the apparent target brightness at the time of the UVES
observations. Table~\ref{charact} lists some basic characteristics of the
final targets. In addition to the bulge objects, a few field AGB stars
that have partly been checked for their content of Tc previously (see
paper~I and references therein) were observed.

The observations with UVES at ESO's Very Large Te\-le\-sco\-pe (VLT),
Cerro Paranal/Chile, were carried out between July~6 and July~9, 2000.
The setting was chosen in order to cover the blue (central
wavelength 4370~\AA) and red (central wavelength 8600~\AA) arm of the
spectrometer simultaneously. The observed wavelength ranges thus were
approximately 3770--4900~\AA\ (blue arm), 6670--8470~\AA\ (red lower
arm), and 8650--9920~\AA\ (red upper arm). With the blue arm, several
resonance lines of neutral Tc were covered, among them the lines at
4238.19, 4262.27 and 4297.06~\AA\ (wavelength in air, hereafter
referred to as the ``classical'' lines). The slit width of the
spectrometer was set to 0\farcs7 which results in a resolution
of $\lambda/\Delta\lambda \cong 50\,000$. The spectra were
taken in the $1 \times 2$ binning mode. Cumulative exposure times
ranged from 600 to 7200~s, with 3600~s as a typical value. All stars
observed during the first night were re-observed in the second or
third night because the exposure times chosen in the first night
were not sufficient.

The spectra were reduced with the ESO provided pipeline written in
MIDAS, version 2.1.0. For the spectra in the blue arm, ``optimal''
extraction was used during the reduction process, whereas ``average''
extraction was used for the red arm. This procedure is recommended by
the pipeline manual to optimise the signal to noise ratio (SNR) of the
spectra. In optimal extraction, the pipeline performs a Gaussian fit
to the signal profile in spatial direction; the SNR is then computed
from the deviation of the profile from this fit.

The achieved SNR of the UVES spectra varies strongly with wavelength.
For the regions around the Tc lines the SNR provided by the UVES
pipeline lies between 5 and 40 with the majority of the stars having
an SNR of 15 or better. The four stars with Tc have an SNR of around
30 (M626, M1347, S942) and 5 (M1147), respectively. However, as
shown below, deciding whether Tc is present is possible at such a
low SNR.

\begin{figure*}
\centering
\includegraphics{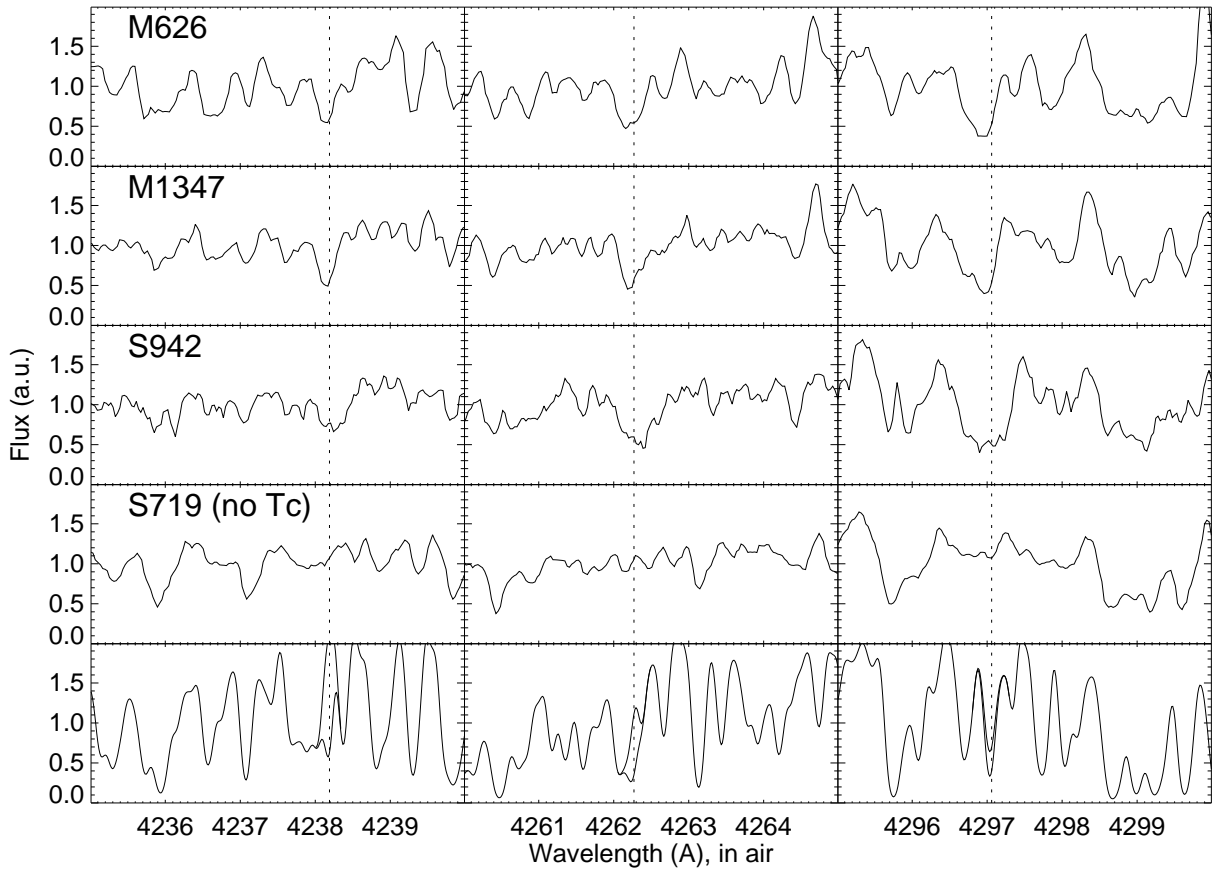}
\caption{Regions around the three classical lines of neutral
Technetium covered by the UVES spectra. The three upper panels show
the stars with definite Tc detection and high signal to noise ratio,
the fourth panel shows a star without Tc. The theoretical positions of
the Tc lines are marked by the dotted vertical line. The synthetic
spectra with and without Tc in the lower panel are based on a
hydrostatic MARCS atmospheric model with T$_{\mathrm{eff}}$ = 3400~K,
[Fe/H] = $-0.5$, $\mathrm{log} (g)$ = 0.0, one solar mass, solar C/O
ratio and a microturbulent velocity $\xi=2.5$~kms$^{-1}$. The spectrum
with Tc was calculated with a Tc abundance of
$\mathrm{log}(\mathrm{A(Tc)}/\mathrm{A(H)}) + 12.0 = 0.0$.
The residual line at the position of the 4297~\AA\ Tc line in the
spectrum calculated without Tc is due to chromium. All spectra are
normalised in order that the mean over the respective plotted region
is 1.0.}\label{TcPlot}
\end{figure*}

\section{Sample characteristics}
\label{SectResults}

Table~\ref{charact} lists some basic characteristics of the observed
targets. The coordinates are the (rounded) positions as found in the
2MASS catalogue \citep{2MASS}.

\begin{table}
\caption{IRAS colours of the objects that could be identified either in
the IRAS Faint Source Catalogue \citep{Moshir} or in the IRAS Point Source
Catalogue \citep{JIRAS}. The zero magnitude fluxes are 28.3~Jy in [12] and
6.73~Jy in [25] \citep{JIRAS88}. The errors are based on the
relative flux uncertainty for the [12] and [25] flux as given in the
respective catalogue, and on the error-bar in K used in Fig.~\ref{FigPK}.
The respective colour value is quoted in brackets if a flux with quality
flag lower than 2 is involved, and no error is given for these cases.}
\label{IRAS}
\begin{tabular}{lcll}
\hline\hline
Name           & IRAS ident.         & K$_0$ - [12]   & [12] - [25]    \\
\hline			   			     	 
M100  & \object{18102-3640} &  2.49$\pm$0.26 &  1.63$\pm$0.23 \\
M315  & \object{18134-3241} &  2.14$\pm$0.14 &  (1.78)        \\
M331  & \object{18136-3207} &  2.13$\pm$0.23 &  (1.92)        \\
M1147 & \object{18297-3624} &  2.36$\pm$0.10 &  0.64$\pm$0.15 \\
M1179 & \object{18305-3503} &  3.32$\pm$0.30 &  0.92$\pm$0.39 \\
M1313 & \object{18342-3415} &  2.92$\pm$0.33 &  0.93$\pm$0.26 \\
M1347 & \object{18354-3436} &  2.11$\pm$0.16 &  0.77$\pm$0.18 \\
S1204 & \object{18313-3402} &  2.51$\pm$0.12 &  (1.96)        \\
\hline
\end{tabular}
\end{table}

The periods in column~4 are taken from \citet{Wess87}. For two
stars (S942 and M315), they give a low quality flag for the period
determination. Using their period measurement (176 and 173.6~d,
respectively), these two objects would be placed outside the range
of the bulge in the period K-magnitude diagram. We searched the
literature and found a published period of 338~d for S942 based on
photometric data \citep{Plaut71}. Note also, that this star was
classified as Mira variable by \citet{Plaut71}. For M315 a light
curve from the MACHO survey could be extracted (as a matter
of fact, this star is the only one in our sample covered by the MACHO
fields). Using the program PERIOD04 \citep{LenzBreger} a robust period
of 326.8~d was determined.

The $J_0$  and $K_0$ magnitudes are the de-reddened mean values of
the ESO, DENIS and 2MASS measurements. We thus have two to five
measurements in both bands for our stars, with two measurements
for only four of the objects (partly due to DENIS values with bad
quality flag). The de-reddening was performed using the linear
relation for the reddening in the $B_J$ band as a function of
galactic latitude as given in \citet{Schulthe98}. To translate
this into the extinction in the J and K band, we use the relation
$R = A_{V}/E(B- V)$ with $R = 3.2$ and the reddening law of
\citet{GS2003}. This results in an extinction of around 0.1 mag in
J and a few 0.01 mag in K.

To cross check our approach we also determined the extinction in
K for the respective object positions using the RGB fitting method,
taking the RGB of 47~Tuc as reference
\citep[following a suggestion by ][]{Mess05,Dutra03}.
We extracted all objects 4\arcmin\ around the target position from
the 2MASS catalogue and constructed a $K_S$ versus $J -K_S$ diagram.
The computed shift relative to the reference RGB resulted in a
small (few 0.01) or even negative extinction value. This proves
that there is no strong, patchy absorption in the direction of any
one of the sample stars, but it also proves that this method is
too insensitive in the outer bulge to reliably determine the
extinction.

The radial velocities in column~7 of Table~\ref{charact} were determined
from the UVES spectra using a cross correlation technique with a
synthetic spectrum as a template. For each of the blue, red lower
and red upper arm wavelength range a region of around 50~\AA\ was
used for the correlation. All three regions were chosen to avoid
very broad absorption and emission lines. For the respective
region in the red lower arm, the TiO band head at 7054 \AA\ was
covered, since this sharp feature results in a very accurate velocity
measurement. The scatter of the derived radial velocities from the
three wavelength regions is of the order of 1~kms$^{-1}$.

Taking a search radius of 30\arcsec, for eight of the sample stars a
corresponding IRAS source could be found \citep[see also][]{Blommaert}.
The identification was checked using the 2MASS K-images. Where
available, the flux from the IRAS Faint Source Catalogue \citep{Moshir}
was taken, otherwise the IRAS Catalogue of Point Sources \citep{JIRAS}
entry was taken, since the relative flux uncertainty quoted is
lower in the former. The IRAS-colours K$_0$--[12] and [12]--[25] are
summarised in Table~\ref{IRAS}. Note that colours were not calculated
using flux ratios but rather zero point magnitudes for each filter
\citep{JIRAS88}. The two stars with Tc in this small sample are the
ones with the longest pulsation period. They have the lowest values
for their IRAS [12]--[25] colour, but are otherwise not conspicuous.
From Fig.~21 of \citet{Whitelock94} and using the K$_0$--[12] colour
we can estimate the mass loss of the IRAS sources in our sample to
be in the range
$-6.8 \lesssim \mathrm{log}(\dot{M}/M_{\sun} yr^{-1}) \lesssim -6.0$.
Selecting optically bright targets for spectroscopy naturally
avoids highly obscured (i.e., high mass loss) objects. We therefore
assume that the ``Tc yes'' stars do not have a considerably higher
mass loss and intrinsic reddening than the other stars in our
sample.

\section{Tc detection}

In order to determine whether or not a star shows Tc we first
inspected the spectra visually around the classical Tc lines
together with synthetic spectra (Fig.~\ref{TcPlot}). The
synthetic spectra are based on MARCS (Model Atmospheres in a
Radiative Convective Scheme) hydrostatic atmospheric
models from \citet{Gustafsson} with improvements introduced
by \citet{Jorgensen}, and with spherical radiative transfer
routines from \citet{Nordlund}. The spectral synthesis was
recently improved by \citet{Gorfer} and atomic line wavelengths
are taken from the VALD data base \citep{Kupka}\footnote{As it
seems not to be documented elsewhere, we note that the
wavelengths of VALD are for vacuum only for wavelengths below
2000~\AA, otherwise for air!}. For Tc, the $gf$-values of
\citet{Bozman} were used. The synthetic spectra were convolved
with a Gaussian to reduce them to a resolution of 50\,000,
matching the resolution of the observed spectra. No
additional macroturbulence was assumed. This assumption was checked
by determining the FWHM of a Gaussian fit to a selection of a few strong,
seemingly unblended lines of Fe, V, Ti and Cr in one of the observed
spectra with high SNR and the generic synthetic spectrum (see below).
The assumption of zero macroturbulence turned out to be acceptable: the
line-widths where never broader by more than 1.2~kms$^{-1}$ in the
observed spectrum than in the model spectrum.

On visual inspection, three stars were identified to display Tc
lines. In Fig.~\ref{TcPlot} we show sections of the observed
spectra around the classical Tc lines of these stars, as well
as of one star without Tc, along with two synthetic spectra
calculated with and without Tc, respectively. The synthetic
spectra are based on a MARCS atmospheric model with
T$_{\mathrm{eff}}$ = 3400~K, [Fe/H] = $-0.5$, $\mathrm{log} (g)$ = 0.0,
one solar mass, solar C/O ratio and a microturbulent velocity
$\xi=2.5$~kms$^{-1}$. The atmospheric parameters
are not meant as a fit to the real objects but rather are the
result of an ``educated guess''. The spectrum with Tc was
calculated assuming a Tc abundance of
$\mathrm{log}(\mathrm{A(Tc)}/\mathrm{A(H)}) + 12.0 = 0.0$.
According to \citet{Schatz}, the equilibrium Tc abundance
resulting from the s-process is 0.37 on this scale.
Admittedly, the model spectra do not fit to the observed
spectra very well as the lines seem to be much more
pronounced in the model spectra than what is observed.
Hence, we do not over-plot them on the observed spectra
but plot them separately in the lower panel of Fig.~\ref{TcPlot}.
The reason for this discrepancy is not
entirely clear. Reducing the metallicity for the model spectra
calculation by -1.5~dex results in comparable line strengths.
Such a low metallicity is unrealistic for stars of an inferred
age of 3~Gyrs and metallicity estimates for the PG3 field
\citep{Schulthe98} slightly above the LMC metallicity. Also, a
higher temperature of the sample stars does not serve as explanation,
since in this case no agreement in the strength of the TiO band heads
would be found, and the $J - K$ colours are incompatible with a
considerably higher temperature.
Various authors \citep[e.g.][]{Merrill62,DominyWallerstein86} reported
on a similar phenomenon observed in field Miras and named --
in the absence of a clear physical explanation -- ``line weakening''.
This effect is apparently not accounted for in the hydrostatic
model atmospheres. Currently, the dynamic model atmospheres of
\citet{Hoefner} are under investigation for their ability to
model atomic line strengths in the blue/visual range.

In Fig.~\ref{TcPlot}, the three stars classified as containing Tc by
visual inspection clearly show the Tc lines. At the position of the
4297~\AA\ Tc line there appears to be a somewhat smaller line also in
the star that was classified as not having Tc (lower panel). This line
is caused by chromium \citep{Little79}. In the three stars identified
to show Tc by visual inspection, also the Tc lines at 3984.97, 4031.63,
4049.11 and 4095.67~\AA\ identified by \citet{Bozman} and listed in the
NIST atomic line database
(http://physics.nist.gov/cgi-bin/AtData/lines\_form) were clearly
identified, whereas they are absent in the other stars.

%
\begin{figure}
\centering
\includegraphics[width=8.5cm]{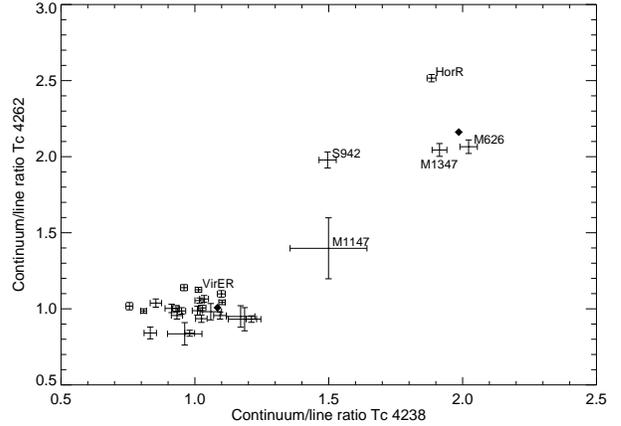}
\caption{Continuum to line flux ratios for the Tc lines at 4238~\AA\
and 4262~\AA. For the former the flux between 4239.4\,--\,4239.7~\AA\
(continuum) was ratio-ed to the flux in the range
4238.0\,--\,4238.3~\AA\ (line), for the latter the wavelength ranges
were 4261.3\,--\,4261.5~\AA\ and 4262.1\,--\,4262.3~\AA, respectively
(see also Fig.~\ref{TcPlot}). Six to ten pixels are typically covered
by these ranges. We include the field stars \object{R~Hor} and
\object{ER~Vir} (analysed in paper~I) for this plot. Also the star
M1147 does separate quite clearly from the compact group of stars
without Tc. This star was not suspected to have Tc from the visual
inspection of its spectrum (due to the low signal to noise ratio). The
filled diamond symbols represent the flux ratios derived from the
synthetic spectra plotted in Fig.~\ref{TcPlot}.}
\label{TcSearch}
\end{figure}
%

For a more quantitative determination of the presence of Tc we define
flux ratios between the Tc line centre and a pseudo continuum point
close to the line position. This continuum point was selected to be
free of any visible absorption features in all observed spectra having
good SNR, and in representative synthetic spectra. In a plot of this
ratio versus the respective ratio for another Tc line, all stars that
were identified to have Tc by visual inspection clearly separate from
the other stars. This was done for several pairs of Tc lines always
giving the same qualitative result. In Fig.~\ref{TcSearch} we show
this ratio for the lines at 4238 and 4262~\AA, as these appear to be
the strongest and least blended lines. The error bars on the flux
ratios were estimated by adding random noise with the magnitude of the
inverse SNR that was provided by the pipeline to the observed
spectrum. The standard deviation in the flux ratio derived from 100
such realisations of the spectrum gives the error bar. M1147, a star
not suspected to show Tc based on the visual inspection, separates
from the compact group of ``Tc no'' stars as well, although not so
obviously. Though this star is the brightest one in the sample in the
K band, its flux in the blue spectral region is so low that the SNR
per pixel around the Tc lines is as low as 5 after one hour of
integration. Taking into account also the flux ratios of the other
identified Tc lines and the error bars from our simulation this star
has to be classified as ``Tc yes'', maybe with a somewhat reduced Tc
abundance with respect to the other ``Tc yes'' stars. For the few
other stars with a low SNR the occurrence of Tc can be definitively
excluded based on these flux ratios. We also include the flux ratio of
the synthetic spectra shown in Fig.~\ref{TcPlot} in this diagram
(filled diamond symbols).

The Tc rich stars are among the coolest and longest-period
in our sample, thus we can exclude that any noteworthy Tc abundance
could have been overlooked in the increasing density of the line forest.
On the other hand, Tc has been detected in field stars significantly
hotter \citep[e.g. $o^1$~Ori,][]{LebzelterHron}. Thus we may
safely assume that within the temperature range of AGB stars the
detection probability is independent of the star's temperature.

We want to briefly summarise here again for using Tc in this work.
Besides Tc, the isotopic ratio $^{12}$C/$^{13}$C certainly is useful as
another indicator of 3DUP. It can be determined from CO lines especially
present in the K band. The advantage of Tc lines as 3DUP indicator is
their complete independence of the star's chemical history due to
the radioactive nature of Tc. This certainly is not the case for the
$^{12}$C/$^{13}$C ratio since it is influenced at least by the initial
isotopic composition, and by the effects of the first (and, in higher
mass stars, the second) dredge-up. Additionally, Tc gives evidence
about recent s-process as well. For comparison, a study on the
$^{12}$C/$^{13}$C ratio in our sample stars and its correlation to the
presence of Tc is under way.

\section{Discussion}

\subsection{Membership in the bulge}

\begin{figure}
\centering
\includegraphics[width=8.5cm]{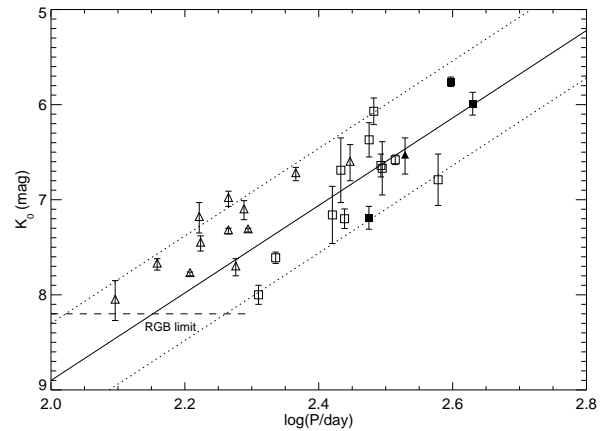}
\caption{Period - K magnitude diagram for our sample of long period
variables in the galactic bulge. Squares and triangles represent Mira
and Semiregular variables, respectively. The filled symbols are stars
with positive Tc detection. The solid line is the period - K relation
from \citet{GS2003}.}
\label{FigPK}
\end{figure}

As basis for the discussion on the bulge membership of the sample
stars, we show a period K-magnitude diagram in Fig.~\ref{FigPK}. The
error-bar in K is the statistical standard deviation of the mean.

We plot the relation from \citet[sequence ``C'']{GS2003} as solid
line instead of the relation from \citet{Glass95} that was used for the
sample selection. The former is an improved version of the latter.
Although \citet{GS2003} note that ``a very few extra observations
could change the slope'', the rather steep slope of \citet{GS2003}, which
is based on single epoch observations in K, fits our multi-epoch
data (two to five measurements in K) quite well. The dotted lines
0\fm6 above and 0\fm5 below sequence ``C'' mark the range in magnitude
due to the finite depth of the bulge \citep{Schulthe98}.
In \citet{GS2003} sequence ``C'' extends in the range
$2.2 < \mathrm{log} P < 2.7$. In the absence of an alternative relation
for the shortest period stars we plot this over the whole diagram. The
dashed line gives the approximate upper limit of the RGB (Tiede, Frogel
\& Terndrup, 1995; Omont et al., 1999; Zoccali et al., 2003). Using this
PL-relation, all stars can be considered to be located in the bulge,
within the error bars.

\citet{GB2005} published a logP-K relation for bulge AGB variables
based on OGLE light curves as well as 2MASS and DENIS data. Since their
linear regression to sequence ``C'' is significantly below the data points
for $\mathrm{log} P > 2.5$ (see their Fig.~3), we do not use their
relation here.
Unfortunately, none of our sample stars is covered by the OGLE survey.
Even using the relation of \citet{GB2005} and adopting the same range
in magnitude for bulge stars as before, only one of the Tc rich stars
(M1147) would be placed in the foreground.
Thus, the conclusion that AGB stars in the bulge with recent dredge-up
are identified is not altered.

Apparently, the SRVs mainly fall above the period K-magnitude relation
in Fig.~\ref{FigPK}. As the PG3 SRVs are in the same pulsation mode
as the Mira variables \citep{Schulthe98}, this fact can be explained
by a selection bias towards brighter, i.e. closer, SRVs, as stars on
the far side of the bulge may fall below the chosen RGB limit.
Detection and non-detection of Tc is marked in this diagram by filled
and open symbols, respectively.

The ``Tc yes'' star with the shortest period is M626 with a period
slightly below 300~d. The two stars with the longest period in our
sample (M1347 and M1147) both show Tc. This compares well with the
findings of paper~I where the fraction of ``Tc yes'' among Mira
variables increases above a period of 300~d. Following the variability
classification of \citet{Plaut71}, only Miras are found to show Tc in
their spectrum (we still keep the SRV symbol for S942 in the
figures).

\subsection{Third dredge-up luminosity limit}

To asses theoretical predictions on the minimum luminosity required
for 3DUP to occur, we constructed a colour-luminosity diagram shown in
the upper panel of Fig.~\ref{MbolJK}. The bolometric magnitudes were
calculated from the K brightness and a bolometric correction based on
$(J - K)$ using the relation of \citet{Kersch06}. This is based on
near infrared and IRAS photometry of a large collection of long period
field variables.
Their relation leads to bolometric magnitudes 0.08 mag fainter on
average than when using the relation of Whitelock, Marang \& Feast
(2000). Furthermore, a distance modulus of 14\fm5 to the bulge was
assumed \citep{McNamara}. 

In paper~I we estimated the minimum luminosity required for 3DUP.
This was derived from the luminosity evolution of a 1.5 M$_{\sun}$
model at the time when 3DUP sets in \citep{Straniero97}. At solar
metallicity, 1.5 M$_{\sun}$ is about the minimum initial mass
required for a star to experience 3DUP on the AGB. This minimum
luminosity corresponds to a bolometric magnitude of
M$_{\mathrm{bol}}=-3\fm9$. In Fig.~\ref{MbolJK} this minimum bolometric
magnitude is drawn as a dotted horizontal line. As can be seen, all
stars with positive Tc detection clearly fall above this line, confirming
the theoretically estimated luminosity limit for 3DUP.

The scatter in magnitude of our sample stars around the log~P-K
relation may be due to various reasons: incomplete light-curve coverage,
depth effects within the bulge, or a scatter in mass and metallicity.
For the lower panel of Fig.~\ref{MbolJK} we assume the scatter to be
solely depth induced (the periods are known with a much higher
precision than the K-magnitudes). To correct for this scatter, we
subtract (or add) the difference between measured K magnitude and the
period K-magnitude relation of \citet{GS2003} from (or to) the bolometric
magnitude. In other words, using the PL-relation we calculate a
distance modulus for every single star. Applying this correction, the
``Tc yes'' stars are the brightest objects at a given $(J - K)_0$
colour.


\subsection{The mass and age of the Tc stars}

From stellar evolution models \citep{Straniero03} one would expect a
minimum initial mass limit for a star to experience 3DUP of 1.4
(Z=0.004) to 1.5~M$_{\sun}$ (solar metallicity). This implies a
limiting age of the ``Tc yes'' stars in our sample of 3 to 4~Gyrs.

Various age estimates of the bulge can be found in the
literature. \citet{Schulthe98} give an age range of 5 to 10~Gyrs from
their study of AGB variables in the PG3 field. \citet{Zoccali}
obtained colour-magnitude diagrams in the visual and near-IR range and
favour a single age of 10~Gyrs, although an age of 5~Gyrs can not be
completely excluded from their analysis (The WFI field studied by
\citet{Zoccali} does not overlap with, but is situated slightly closer to
the galactic plane then the PG3 field). \citet{GB2005} studied the Mira
population in the OGLE bulge fields and derive an age of 1~--~3~Gyrs. Also
\citet{Zoccali} find a number of stars significantly brighter than the
estimated RGB tip but do not interpret this as a sign for an
intermediate age population. Studies on the inner part of the
bulge \citep{vanLoon} and of the galactic bar \citep{ColeWeinberg} find
signatures of an intermediate age (1 to several Gyrs) population on top
of the main old component.


An interesting age indicator may come from the period distribution
of the Miras in our sample. As argued by \citet{HughesWood}, the period
distribution of Miras found in the LMC can be understood as a combination
of an intermediate and an old population among these variables. Short
period Miras (around 200~d) are thought to be older and more metal poor
than their long period counterparts \citep{Hron91}. As our sample also
includes both short and long period Miras, we may suspect in analogy to
the LMC and the galactic field that the bulge contains stars of a
considerable age range. \citet{GB2005} also conclude from the period
distribution of the Mira stars in the OGLE fields that the long period
stars must originate from an intermediate age population.

\begin{figure}
\centering
\includegraphics[width=8.5cm]{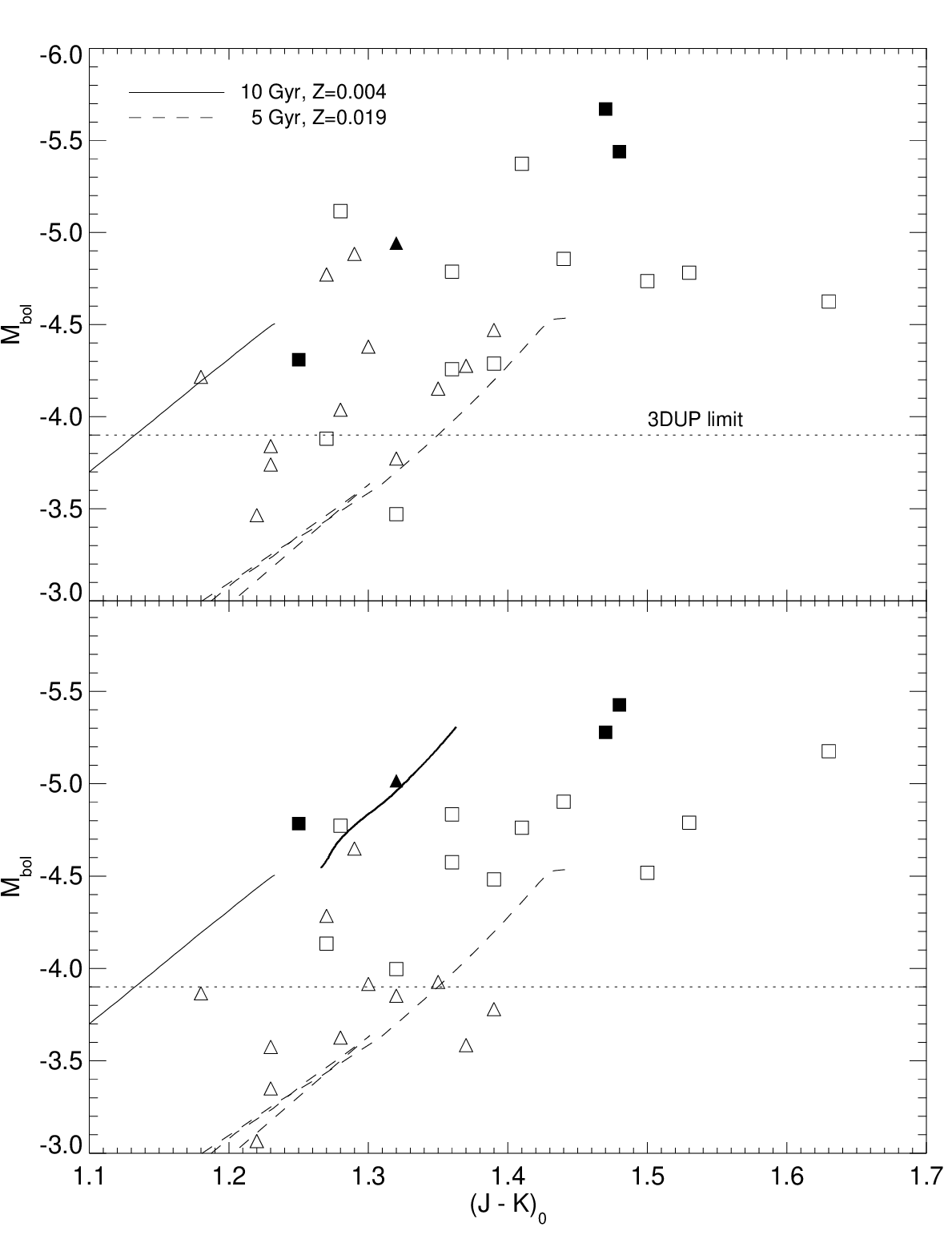}
\caption{Colour-luminosity diagram of the sample stars; symbols are the
same as in Fig.~\ref{FigPK}. In the upper panel, the luminosities as
directly derived from the near infrared photometry are plotted, whereas in
the lower panel the luminosity is corrected for the depth induced scatter
using Fig.~\ref{FigPK}. The dotted horizontal line marks the minimum
luminosity at the stage where 3DUP sets in. Isochrones from
\citet{Girardi2000} represented in solid and dashed lines are also included
in the plot (see legend). For both isochrones the tip of the AGB is well
below the observed AGB tip. The thick line in the lower panel is a
1.5~M$_{\sun}$ evolutionary model track from Straniero.}
\label{MbolJK}
\end{figure}

In Fig.~\ref{MbolJK} we include two isochrones from \citet{Girardi2000}
with two age metallicity combinations (see legend). The original
$(J - K)$ colours of \citet{Girardi2000} do not reach values above 1.3
and thus do not bracket the observational data properly. This is
probably related to the fact that the colours where derived from
hydrostatic model atmospheres while most of the red luminous stars are
strongly pulsating objects with dynamic atmospheres. Therefore we chose
an observational approach and combined the effective temperatures of
\citet{Girardi2000} with a  $(J - K)$ vs. $T_{\mathrm{eff}}$
calibration determined from interferometric data and near infrared
photometry of field Mira variables \citep[see][]{Schulthe98}. With our
calibration the isochrones reach to redder $(J - K)$ colours, thus becoming
a more realistic description of the observations. But still, the isochrones
do not cover the reddest objects, and both parameter sets have their AGB
tip at a luminosity fainter by one magnitude than what is observed. The
picture is not changed in the second version (lower panel) of the diagram.
The extension to red colours illustrates the general difficulty of
transforming the effective temperature of a static evolutionary model
into colours of a strongly pulsating atmosphere and a systematic
investigation of this point is desirable. The problem of the tip luminosity
is discussed in the following.

Taking the isochrone of 2.82~Gyrs age of \citet{Girardi2000} leads to an
AGB tip luminosity of only M$_{\mathrm{bol}}=-4\fm8$. Taking into account
the luminosity variation during a thermal pulse cycle (see next paragraph)
thus slightly underestimates the AGB tip luminosity. This implies that models
with large overshoot from the envelope, like the ones of \citet{Girardi2000},
lead to luminosities somewhat lower than what is observed.


In the lower panel of Fig.~\ref{MbolJK} (corrected for the depth
scatter) we include an evolutionary model track from O. Straniero (priv. comm.)
as a thick solid line. The model used here is similar to the ones presented in
\citet{Straniero97}, only that in the current model mass loss is taken into
account. The same $(J - K)$ vs. $T_{\mathrm{eff}}$ calibration
as for the isochrones was used to include the model track into this diagram.
The parameters of the model are as follows: initial mass 1.5~M$_{\sun}$,
metallicity $Z = 0.02$, Helium fraction $Y = 0.28$, and Reimers mass loss
parameter $\eta = 0.5$ \citep{Reimers}. The track covers only the evolution
between the pre-last and the last TP of this model, excluding the short term
luminosity spikes at the TPs themselves (the luminosity in between the
preceding TPs is only slightly lower). The track spans about 60\,000 years.
The evolution runs from the lower left end to the upper right end of the
track. With other words, in-between two TPs the luminosity increases while
the temperature drops, until the next TP ``resets'' the star back to low
luminosity and high temperature. The bolometric magnitude varies by 0\fm75
between two TPs, whereas the temperature varies only slightly between 3176
and 3112~K.

The luminosities of the evolutionary model track match those of the
``Tc yes'' stars quite well, especially the tip luminosity is in remarkable
agreement. Only the range in $(J - K)$ of the observed stars is larger than
that of the model track, either because the ``Tc yes'' stars span a bigger
range in temperature or because of the mentioned problems involved in the
$T_{\mathrm{eff}}$ to $(J - K)$ transformation. In the light of these results
it is thus not surprising to find stars more luminous than the tip of the
Girardi isochrones. All ``Tc yes'' stars are found there suggesting that
these stars are indeed the most evolved and most massive ones in our sample.

Besides the evolutionary track of Straniero also the
interpolation formulae of \citet{Straniero03} were used for a comparison
with the present data set. With these formulae, the luminosity and the
temperature at half the time between two successive TPs can be calculated.
Using the same parameters as for the full model track very good agreement
with the observational data could be found as well.

Trusting in the nucleosynthesis and stellar evolution models, our results
would require that the bulge includes a population of stars with an age
around 3~Gyrs. On the other hand, if the age estimates of \citet{Schulthe98}
and \citet{Zoccali} are correct, then dredge-up would occur at a minimum
mass significantly lower than 1.4~M$_{\sun}$. This would further imply that
the maximum AGB luminosity predicted by the stellar evolution models like
the ones from \citet{Straniero03} is too low.

%

\begin{figure}
\centering
\includegraphics[width=8.5cm]{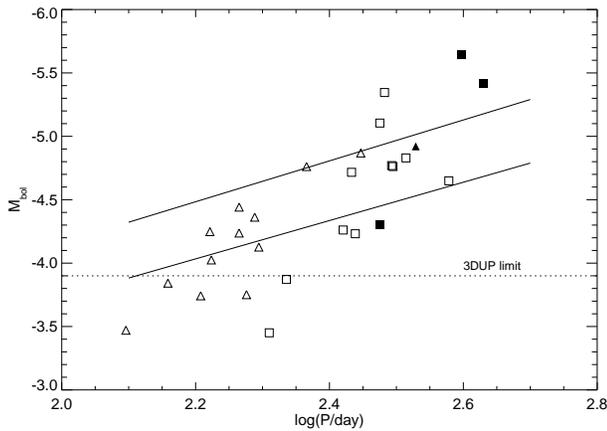}
\caption{Bolometric magnitude versus period of the sample stars.
Again, the symbols are the same as in the previous figures.
The dotted horizontal line marks the minimum bolometric magnitude
at the stage where 3DUP sets in. The solid lines are relations
for a 1~M$_{\sun}$ (lower line) and a 1.5~M$_{\sun}$ (upper line)
model star pulsating in fundamental mode \citep{WoodSebo96}.}
\label{MbolP}
\end{figure}

Finally, in Fig.~\ref{MbolP} we present a M$_{\mathrm{bol}}$ log($P$)
diagram of our sample stars. The distribution of the objects is very
similar to the one in Fig.~\ref{FigPK}. The solid lines drawn are
theoretical relations from \citet{WoodSebo96} for a 1~M$_{\sun}$
(lower line) and a 1.5~M$_{\sun}$ model (upper line) in fundamental
mode pulsation. Again, as for the evolution models, masses around
1.5~M$_{\sun}$ are required to fit the brighter, long period pulsators
in the present sample. This supports the solution discussed above that
there is a young, more massive population present in the outer bulge
and that the predictions of mixing theories are correct.


%

\section{Conclusions and Outlook}

We presented high resolution UVES/VLT spectra and NIR photometry
of bulge AGB variables, aiming at a detection of third dredge-up
indicators, namely Technetium. The bulge membership of these stars
is discussed using a period K-magnitude diagram. In a sample of
27 stars, four were found to  have Tc, giving the first direct
evidence for recent or ongoing third dredge-up in these stars.
For the distinction between ``Tc no'' and ``Tc yes'' stars, we
compare the observed spectra to synthetic ones around the
``classical'' Tc lines. A more precise distinction is possible
using flux ratios between the Tc line flux and a pseudo-continuum
flux. Using this method, even for very low signal to noise ratio
spectra (down to 5) a reliable distinction of this kind can be
made.

In a colour luminosity diagram of the sample stars all objects with
Tc clearly fall above the theoretical third dredge-up limit of
M$_{\mathrm{bol}} = -3\fm9$, in agreement with model predictions.

%

Many stars above the theoretical luminosity limit for third
dredge-up do not show Tc. There have been suggestions that
Tc could decay (even below the detection limit) if 3DUP does
not occur for several TPs \citep[see][and paper~I]{Busso92}.
%
One may speculate that the star M1147 has a reduced Tc
abundance with respect to the other three ``Tc yes'' stars,
although it is one of the brightest (and most evolved) objects in the
present sample. It is possible that this star on the AGB tip has
lost so much mass that the envelope is not massive enough anymore to
drive dredge-up, and Tc has already decayed to a lower
abundance level.

The disagreement between observed spectra and synthetic
spectra based on hydrostatic atmospheric models regarding atomic line
strengths in the blue/visual range is a striking phenomenon found for
pulsating AGB variables. In the near future, dynamic models such as
from \citet{Hoefner} will be tested for their ability to reproduce the
observed line strengths.

From the period distribution, the period luminosity diagram and the
detection of Tc as incontestable indicator for 3DUP, a mass of about
1.5~M$_{\sun}$ for at least some of the sample stars is required. This
implies an upper age limit of around 3~Gyrs for these stars,
consistent with other findings of an intermediate age population in
the bulge \citep{vanLoon,GB2005}. Contrary to this, \citet{Zoccali}
does not find any signatures of an intermediate age population but
favours a single age of 10~Gyrs for the bulge. A solution to this
disagreement can not be given here and remains for future work on
the galactic bulge. Isochrones assuming large overshoot from the
convective envelope are found to somewhat underestimate the AGB tip
luminosity, though a younger age of the population would reduce the
discrepancy.

\begin{acknowledgements}
We wish to thank B. Aringer and M. Gorfer for improvements in the
spectral synthesis calculation and including Tc line data.

We also thank M. Messineo (ESO Garching) for inspiring and helpful discussion,
and O. Straniero (Teramo, Italy) for providing the AGB evolution models used
here.

The constructive comments of the referee (M. Reyniers) were very helpful
for improving the paper.

TL acknowledges funding by the Austrian Science Fund FWF under project
P18171.

This publication makes use of data products from the Two Micron All
Sky Survey, which is a joint project of the Uni\-ver\-si\-ty of
Mas\-sachu\-set\-ts and the Infra\-red Pro\-ces\-sing and Ana\-ly\-sis
Cen\-ter/Ca\-li\-for\-nia In\-sti\-tu\-te of Tech\-no\-lo\-gy, funded
by the Na\-tio\-nal Aero\-nau\-tics and Space Ad\-mini\-stra\-tion and
the Na\-tio\-nal Science Foundation.

This research has made use of the SIMBAD database, operated at CDS,
Strasbourg, France.

\end{acknowledgements}


\begin{thebibliography}{}

\bibitem[Abia et al., 2002]{Abia02} Abia C., Dom\'{i}nguez I., Gallino R.,
et al., 2002, ApJ 579, 817

%
%
\bibitem[Barnbaum \& Morris, 1993]{BarnbaumMorris93} Barnbaum C. \& Morris M.,
1993, BAAS 182, 46.17

\bibitem[Blommaert, 1992]{Blommaert} Blommaert J. A. D. L., 1992, Ph.D. thesis,
Leiden University, the Netherlands

\bibitem[Bozman et al., 1968]{Bozman} Bozman W. R., Corliss C. H.
\& Tech J. L., 1968, Journal of Research Nat. Bur. Stand. Vol. 72A, No.6, 559
%

\bibitem[Busso et al., 1992]{Busso92} Busso M., Lambert D. L.,
Beglio L., et al., 1992, ApJ 399, 218

\bibitem[Busso et al., 1999]{Busso99} Busso M., Gallino R. \& Wasserburg G. J.,
1999, ARA\&A 37, 239

\bibitem[Busso et al., 2001]{Busso01} Busso M., Gallino R., Lambert D. L.,
Travaglio C., \& Smith V. V., 2001, ApJ 557, 802

\bibitem[Cole \& Weinberg, 2002]{ColeWeinberg} Cole A. A. \& Weinberg, 2002,
ApJ 574, L43

\bibitem[Costa \& Frogel, 1996]{CostaFrogel} Costa E. \& Frogel J. A., 1996,
AJ 112, 2607

\bibitem[Cutri et al., 2003]{2MASS} Cutri R., Skrutskie M., Van Dyk S. et al.,
2003, The Two Micron All Sky Survey Catalogue of Point Sources

\bibitem[Dekker et al., 2000]{Dekker} Dekker H., D'Odoricio S., Kaufer A.
et al., 2000, Proc. SPIE Vol. 4008, p. 534-545

\bibitem[Deroo et al., 2005]{Deroo} Deroo P., Reyniers M., van Winckel H.
et al., 2005, A\&A 438, 987

\bibitem[Dutra et al., 2003]{Dutra03} Dutra C. M., Santiago B. X.,
Bica E. L. D. \& Barbuy B., 2003, MNRAS 338, 253

\bibitem[Dominy \& Wallerstein, 1986]{DominyWallerstein86} Dominy J. F. \&
Wallerstein G., 1986, ApJ 310, 371

\bibitem[Epchtein et al., 1997]{Epchtein97} Epchtein N., de Batz B.,
Capoani L., et al., 1997, ESO Mes\-sen\-ger 87, 27

%
\bibitem[Girardi et al., 2000]{Girardi2000} Girardi L., Bressan A.,
Bertelli G. \& Chiosi C., 2000, A\&AS 141, 371

\bibitem[Glass et al., 1995]{Glass95} Glass I. S., Whithelock P. A., Catchpole
R. M. \& Feast M. W., 1995, MNRAS 273, 383

\bibitem[Glass \& Schultheis, 2003]{GS2003} Glass I.S. \& Schultheis M., 2003,
Mon. Not. R. Astron. Soc. 345, 39 - 48

\bibitem[Gorfer, 2005]{Gorfer} Gorfer M., Diploma thesis, University
of Vienna, Austria

\bibitem[Goriely \& Mowlavi, 2000]{Goriely} Goriely S. \& Mowlavi N., 2000,
A\&A 362, 599

\bibitem[Groenewegen \& de Jong, 1993]{GroenewegendeJong} Groenewegen M. A. T.
\& de Jong T, 1993, A\&A 267, 410

\bibitem[Groenewegen \& Blommaert, 2005]{GB2005} Groenewegen M. A. T. \&
Blommaert J. A. D. L., 2005, A\&A 443, 143

\bibitem[Guandalini et al., 2006]{Guandalini2006} Guandalini R., Busso M.,
Ciprini S. et al., 2006, A\&A 445, 1069

\bibitem[Gustafsson et al., 1975]{Gustafsson} Gustafsson B., Bell R. A.,
Eriksson K. \& Nordlund A., 1975, A\&A 42, 407

\bibitem[Herwig, 2005]{Herwig} Herwig F., 2005, ARAA 43, 435

\bibitem[H\"ofner et al., 2003]{Hoefner} H\"ofner S., Gautschy-Loidl R.,
Aringer B. \& J\o rgensen U. G., 2003,  A\&A 399, 589

\bibitem[Hron, 1991]{Hron91} Hron J., 1991, A\&A 252, 583

\bibitem[Hughes \& Wood, 1990]{HughesWood} Hughes S. M. G. \& Wood P. R., 1990,
AJ 99, 784

\bibitem[Joint IRAS Science W.G., 1988]{JIRAS88} Joint IRAS Science Working
Group, 1988, IRAS Catalogues and Atlases, Volume 1: Explanatory Supplement,
NASA RP-1190

\bibitem[Joint IRAS Science W.G., 1994]{JIRAS} Joint IRAS Science Working
Group, 1994, IRAS catalogue of Point Sources, Version 2.0

\bibitem[J\o rgensen et al., 1992]{Jorgensen} J\o rgensen U. G., Johnson H. R.
\& Nordlund A., 1992, ASPC 26, 540

\bibitem[Kerschbaum et al., in preparation]{Kersch06} Kerschbaum et al., 2006,
in preparation

\bibitem[Kupka et al., 1999]{Kupka} Kupka F., Piskunov N., Ryabchikova T. A.,
Stempels H. C. \& Weiss W. W., 1999, A\&AS 138, 119

\bibitem[Lattanzio, 2002]{Lattanzio02} Lattanzio J. C., 2002, New Astron. Rev.,
46, 469

\bibitem[Lebzelter \& Hron, 1999]{LebzelterHron} Lebzelter T. \& Hron J., 1999,
A\&A 351, 533 - 542

\bibitem[Lebzelter \& Hron, 2003]{paperI} Lebzelter T. \& Hron J., 2003,
A\&A 411, 533 - 542 (paper I)

\bibitem[Lenz \& Breger, 2004]{LenzBreger} Lenz P. \& Breger M., 2004,
IAU Symposium no. 224, Cam\-bri\-dge University Press, p. 786

\bibitem[Little et al., 1987]{Little87} Little S. J., Little-Marenin I. R.
\& Hagen-Bauer W., 1987, AJ 94, 981

\bibitem[Little-Marenin \& Little, 1979]{Little79} Little-Marenin I.
\& Little S. J., 1979, AJ 84, 1374

\bibitem[Lugaro et al., 2003]{Lugaro03} Lugaro M., Herwig F., Lattanzio J. C.,
Gallino R., \& Straniero O., 2003, ApJ 586, 1305

\bibitem[Marigo et al., 1996]{Marigo96} Marigo P., Bressan A. \& Chiosi C,
1996, A\&A 313, 545

\bibitem[Masseron et al., 2006]{Masseron} Masseron T., Van Eck S., Famaey B.
et al., 2006, astro-ph/0605658, A\&A, in press

\bibitem[McNamara et al., 2000]{McNamara} McNamara D. H., Madsen J. B.,
Barnes J. \& Ericksen B. F., 2000, PASP 112, 202

\bibitem[Merrill, 1952]{Merrill} Merrill P. W., 1952, ApJ 116, 21

\bibitem[Merrill et al., 1962]{Merrill62} Merrill P. W., Deutsch A. J. \&
Keenan P. C., 1962, ApJ 136, 21

\bibitem[Messineo et al., 2005]{Mess05} Messineo M., Habing H. J.,
Menten K. M., et al., 2005, A\&A 435, 575

%
\bibitem[Moshir et al., 1989]{Moshir} Moshir M. et al., 1989, IRAS Faint
Source Sur\-vey, In\-fra\-red Pro\-ces\-sing and An\-al\-ys\-is Cen\-ter,
Ca\-li\-for\-nia In\-sti\-tu\-te of Tech\-no\-lo\-gy, Pa\-sa\-de\-na

\bibitem[Ng, 1994]{Ng} Ng Y. K., 1994, Ph.D. thesis, Lei\-den Uni\-ver\-si\-ty,
the Ne\-ther\-lands

\bibitem[Nordlund, 1984]{Nordlund} Nordlund \AA., 1984, in Methods in Radiative
Transfer, ed. W. Kalkofen (Cambridge University Press, Cambridge), 211

\bibitem[Omont et al., 1999]{Omont99} Omont A. et al., 1999 A\&A 348, 755

\bibitem[Plaut, 1971]{Plaut71} Plaut L., 1971, A\&AS 4, 75

\bibitem[Reimers, 1975]{Reimers} Reimers D., 1975, Mem. Soc. Roy. Sci.
Li\`{e}ege, 6th Ser., 8

\bibitem[Robin et al., 2003]{Robin03} Robin A. C., Reyl\'{e} C.,
Derri\`{e}re S. \& Picaud S., 2003, A\&A 409, 523

\bibitem[Schatz, 1983]{Schatz} Schatz G., 1983, A\&A 122, 327

\bibitem[Schultheis, 1998]{SchultheDiss} Schultheis M., 1998,
Ph.D. thesis, Uni\-ver\-si\-ty of Vien\-na, Aus\-tria

\bibitem[Schultheis et al., 1998]{Schulthe98} Schultheis M., Ng Y. K., Hron J.,
\& Kerschbaum F., 1998, A\&A 338, 581 - 591

\bibitem[Smith \& Lambert, 1988]{SmithLambert88} Smith V. V. \& Lambert D. L.,
1988, ApJ 333, 219

\bibitem[Smith \& Lambert, 1990]{SmithLambert90} Smith V. V. \& Lambert D. L.,
1988, ApJS 72, 387

\bibitem[Straniero et al., 1997]{Straniero97} Straniero O., Chieffi A.,
Limongi M., et al., 1997, ApJ 478, 332

\bibitem[Straniero et al., 2003]{Straniero03} Straniero O., Dom\'{i}nguez I.,
Cristallo S. \& Gallino R., 2003, PASA 20, 389

\bibitem[Tiede, Frogel \& Terndrup, 1995]{Tiede95} Tiede G. P., Frogel
J. A. \& Terndrup D. M., 1995, AJ 110, 2788 

\bibitem[Van Eck \& Jorissen, 1999]{VanEckJorissen99} Van Eck S.
\& Jorissen A., 1999, A\&A 345, 127

\bibitem[van Loon et al., 2003]{vanLoon} van Loon J. T., Gilmore G. F.,
Omont A. et al., 2003, MNRAS 338, 857

\bibitem[Vanture et al., 1991]{Vanture91} Vanture A. D., Wallerstein G.,
Brown J. A. \& Bazan G., 1991, ApJ 381, 278

\bibitem[Wallerstein et al., 1997]{Wallerstein97} Wallerstein G., Iben I.,
Parker P., et al., 1997, Rev. Mod. Phys. 69, 995

\bibitem[Wallerstein \& Dominy, 1988]{WallersteinDominy88}
Wallerstein G. \& Dominy J. F., 1988, ApJ 330, 937

\bibitem[Wesselink, 1987]{Wess87} Wesselink Th. J. H., 1987,
Ph.D. thesis, Catholic University of Nijmengen, the Netherlands

\bibitem[Whitelock et al., 1994]{Whitelock94} Whitelock P. A., Menzies J.,
Feast M. et al., 1994, MNRAS 267, 711

\bibitem[Whitelock, Marang \& Feast, 2000]{Whitelock00} Whitelock P. A,
Marang F., Feast M. W., 2000, MNRAS 319, 728

\bibitem[Wood \& Sebo, 1996]{WoodSebo96}
Wood P. R. \& Sebo K. M., 1996, MNRAS 282, 958

\bibitem[Zoccali et al., 2003]{Zoccali} Zoccali M., Renzini A., Ortolani S.
et al., 2003, A\&A 399, 931

\end{thebibliography}
\end{document}